\documentclass[11pt]{article}
\usepackage{graphicx,url,epstopdf,amsmath,amssymb}

\begin{document}
\title{A reproduction rate which perfectly fits Covid-19} 
\author{Christoph Bandt\thanks{Institute of Mathematics, University of Greifswald, 17487 Greifswald, Germany, bandt@uni-greifswald.de} } 
\maketitle

\begin{abstract}
We present a simple technique to compare the development of the Covid-19 epidemic in different regions, based only on the time series of confirmed cases. Weekly new infections, taken for every day, are interpreted as infection potential of Covid-19. We derive a robust time-varying reproduction rate for the infection potential, including asymptomatic cases, which does not depend on death rate or testing intensity. It requires few assumptions and shows a more plausible time course than official reproduction rates in several countries. 
\end{abstract}

\begin{figure}[h!t] 
\begin{center}
\includegraphics[width=0.48\textwidth]{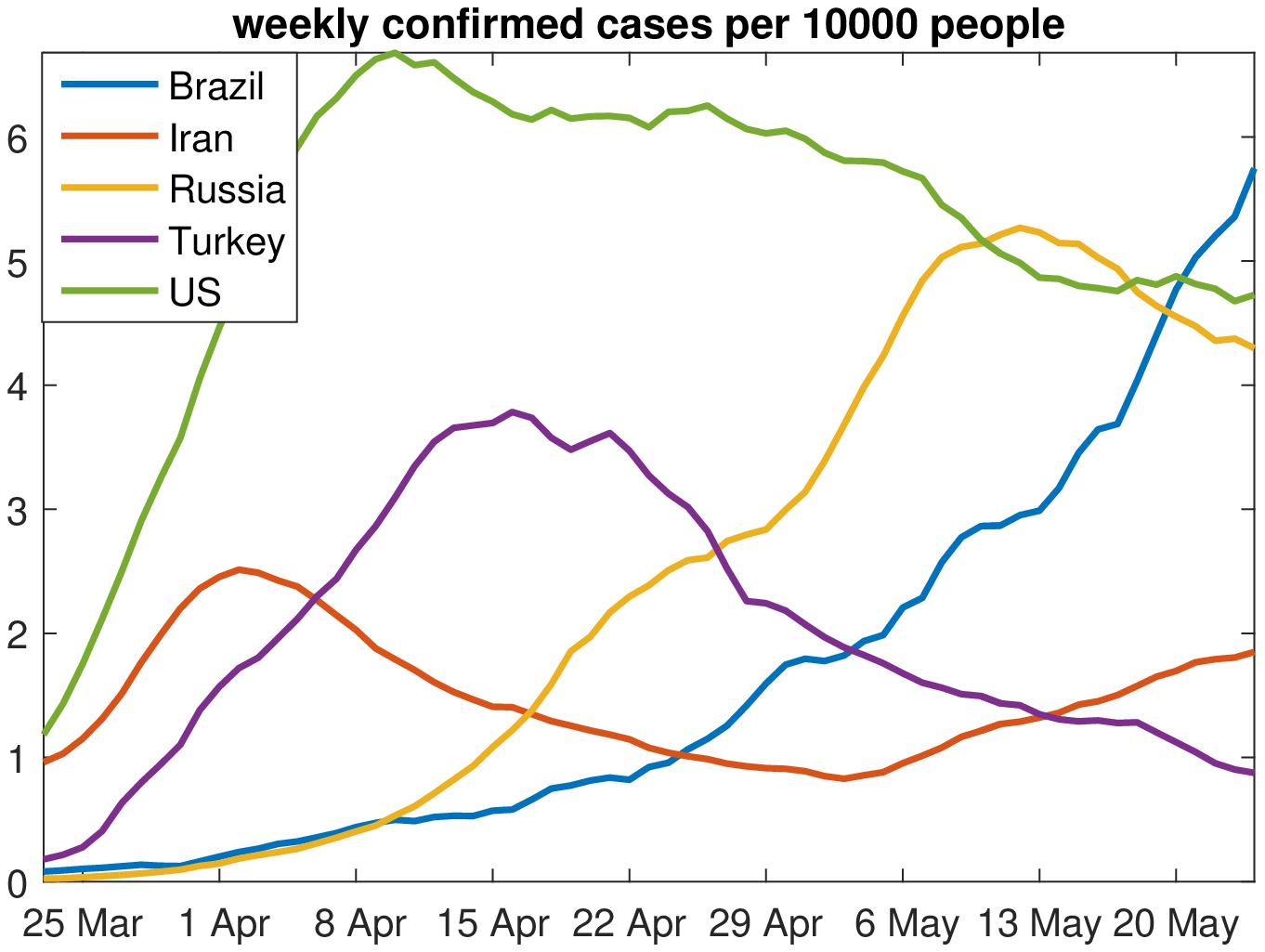}\quad
\includegraphics[width=0.48\textwidth]{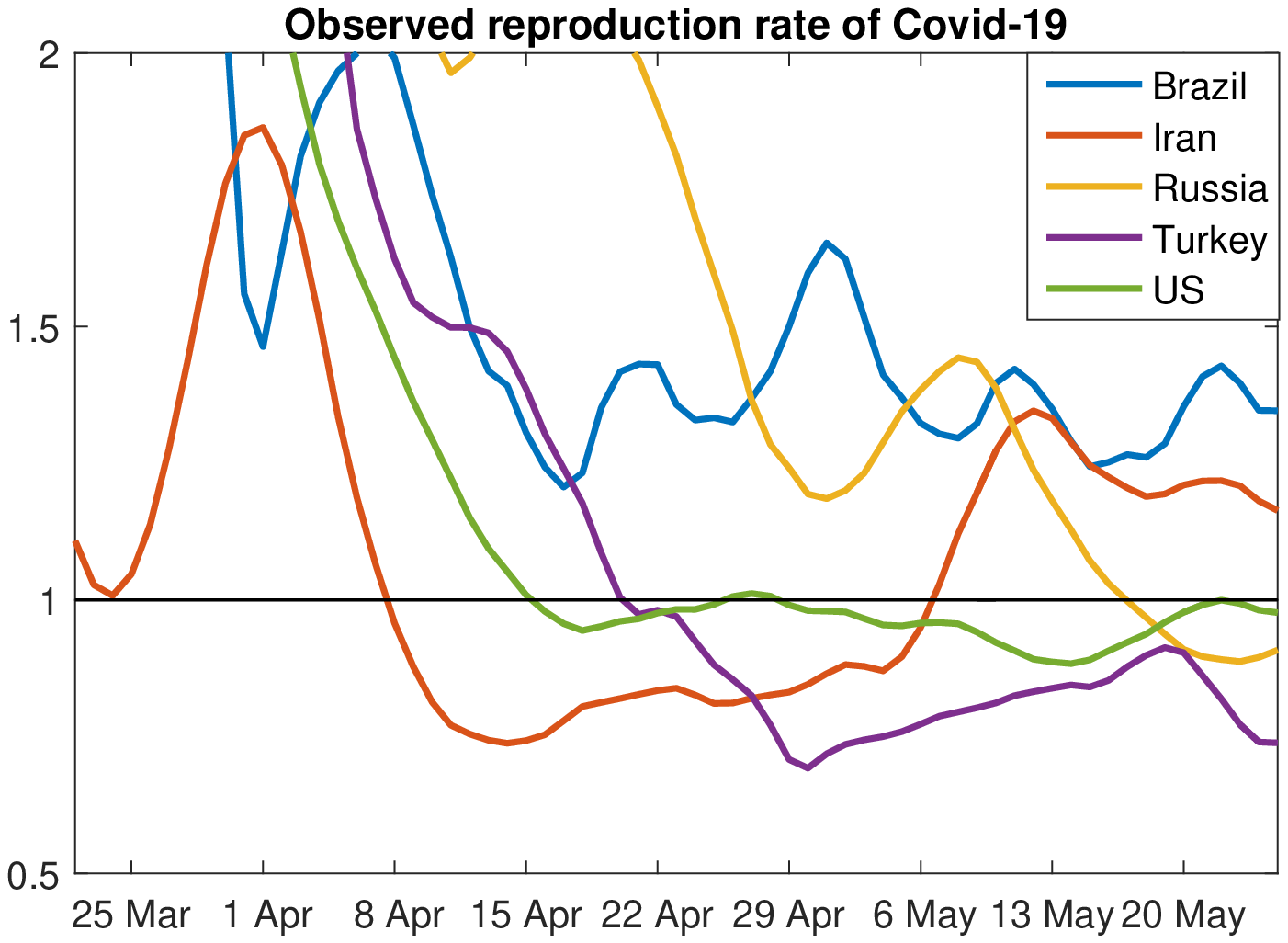}  
\end{center}
\caption{Weekly cases per capita (left) and reproduction numbers (right) for the United States, Russia, Brazil, Iran and Turkey from March 23 up to May 24, 2020.}\label{fj1}
\end{figure}

\subparagraph{Daily and weekly Corona numbers.}  The Covid-19 pandemic is commonly described by cumulated numbers of confirmed cases, deaths and recoveries, for each day $t$ and each country or region. We use only the series $C_t$ of confirmed cases from the database \cite{JHU} of Johns Hopkins University in its version from 25 May 2020. The letter $t$ denotes a date. If $t$ is today, then $t-1$ is yesterday, and $t-7$ is one week ago. $C_t$ is a function of time with a meaning in the first stage of an epidemic, but not for lockdown conditions. The numbers of new infections $N_t= C_t-C_{t-1}$ are the original data collected each day. They show tremendous variation, including a periodic component reflecting the weekly rhythm of health administrations \cite{JHU, Bandt}. 
The best way to represent the course of the epidemic is by weekly new infections 
\begin{equation}
 W_t= C_t-C_{t-7}=N_t+N_{t-1}+...+N_{t-6}  
\label{weekly}\end{equation}
for each day $t$ \cite{OWID, Bandt}, called Covid-19 activity in the maps of \cite{RKI}. In Figure \ref{fj1} they are shown for five countries as incidences, to adjust for population size.

\subparagraph{Three stages of the epidemic, and the delay principle.} 
In a first stage, the epidemic in a country grows exponentially. Weak countermeasures reduce the rate of growth, but will not stop the growth \cite{rep9}. Brazil is still in this phase. The other countries have implemented strong lockdown measures and managed to stop the growth. 

When dealing with Covid-19, we have to distinguish infection time, symptom time, and observation time. \emph{From the time of infection up to the observation of a case in the statistics, there is a delay of 2 up to 3 weeks} \cite{ecdc,RKI}.
The time between infection and onset of symptoms is the incubation period which man cannot control. It is between 2 and 14 days, with a mean of about 5 days \cite{worldo}. 
The other part of the delay includes patient's hesitation and visit to the doctor, testing, laboratory work, and report of the case from local officials up to the government. This time usually includes one or two weekends.

At the lockdown day the delay principle implies a bad prediction: we remain in stage 1 with increasing case numbers for 2 or 3 weeks. Even if the lockdown immediately stops all infections, it cannot avoid the infections which took place before. That is a painful long waiting time for people, politicians and the media who expect immediate relief. Stronger measures are often implemented to demonstrate activity against the pandemic.

After successful containment, and 2-3 weeks waiting time, the apex of infections is reached and a second stage of decline of infections starts. For Iran and Turkey in Figure \ref{fj1}, the initial decline was almost as fast as the previous increase, for the USA it started slowly. Small regions have a good chance to get straight to a low level. Large countries often slow down on intermediate level so that lockdown measures are extended. Due to the delay principle, lifting of containment measures can influence the statistics only 2-3 weeks later. So there is a good prediction for at least 2 weeks.  

In a third stage, the goal is to keep the level of infections low and avoid new waves.  In Figure \ref{fj1}, Iran is in the third phase, opposing a second wave. Asian countries like Vietnam, Thailand, South Korea and Taiwan are on an extremely low level of infections. European countries and the USA can hardly reach that level without severe economic losses. Some Eastern European countries, Norway and Austria presently have the best prospects.

\subparagraph{The difference between infections and confirmed cases.}
Most infections by Covid-19 are caused by people who have little or no symptoms, or no symptoms yet. The number $N_t$ of observed new positive cases at day $t$ is smaller than the unknown number $I_t$ of new infected persons for day $t :$ 
\begin{equation}  I_t = c\cdot N_t \mbox{ for some constant } c>1. 
\label{ccc}\end{equation}
The parameter $c$ is unknown and hard to estimate. According to \cite{Ioannidis} it can be anywhere between 3 and 300. Note that $c$ differs from country to country. In the USA a lot of testing was performed, so $N_t$ includes many mild cases.  Certainly $c$ is much higher for Brazil where only severe cases were tested. Thus  the level of infection in Figure \ref{fj1}, taken as weekly incidence of confirmed cases, is not comparable between countries without further adjustment.  The shape of curves is comparable, however, and indicates the stage of the epidemic. 

On the right of Figure \ref{fj1} there is a function which does not depend on the intensity of testing and allows comparison of different regions. 

\subparagraph{The concept of reproduction rate.}
The reproduction rate $R$ is a central concept in epidemiology \cite{Cori,Ferretti,Aronson,rep21,travel}. It is `the average number' of people who are infected by one sick person. If $R>1,$ the number of infections will increase exponentially. If $R<1,$ the number of infections will decline. In Figure \ref{fj1}, Brazil and Iran have reproduction numbers above 1. Their case numbers are continuing to rise in the coming days.  Russia just managed to get below 1, which means that case numbers are going to decrease further. Turkey and the USA have subcritical reproduction numbers near to one, so it is not clear how long the decline of case numbers will continue.

Case numbers and reproduction rates in
Figure \ref{fj1} must be studied together. Supercritical reproduction on high level of infections will lead to disaster while on low level of infections it can be tolerated for some time. Subcritical reproduction on high infection level can lead to immediate relief while on low level it will not be felt so much. On the whole, reproduction rate is a good concept for large infection numbers. When new infections can be counted with fingers, $R$ becomes obsolete.

Note that reproduction rate, like life expectancy, is an abstract notion and not a statistical average. It is calculated from a variety of models \cite{epibul,Cori,rep21,ages}, often with differing results. Most models require estimation of model parameters. The value of $R$ then comes with a confidence interval indicating its accuracy. The basic model assumptions are taken for granted, however.

Our approach is non-parametric, with few assumptions explained below.  We claim  that the assumptions fit the properties of Covid-19 and of the given data extremely well.

\subparagraph{The infection potential.} 
The following two sections are theoretic. We derive a formula for $R_t,$ where $t$ \emph{now denotes symptom time} - the day of onset of symptoms.
Now $N_t$ will denote the number of sick persons with symptoms day $t$ which will later be confirmed by a test and enter the statistics. 

\emph{Our main assumption is that a sick person remains infectious for 7 days.} More specifically, these are the two days before noting symptoms and four days after. For asymptomatic cases there should also be a `symptom day' which separates the infectious period into two and four days. Let $I_t$ be the unknown number of all symptomatic and asymptomatic people with `symptom day' $t.$ We assume that equation \eqref{ccc} holds for this definition of $N_t$ and $I_t.$
We assume that $c$ can be considered as a constant for a period of 10 days. 

Now we consider the \emph{infection potential} at time $t,$ that is, all asymptomatic or symptomatic persons which are infectious at day $t.$ This includes $I_t$ and the infectious people with symptom day 1 or 2 days after $t$ or up to 4 days before $t.$ So the infection potential at time $t$ is
\begin{equation} 
I_{t-4}+I_{t-3}+...+I_{t+2}=c\cdot (N_{t-4}+N_{t-3}+...+N_{t+2})
= c\cdot W_{t+2}\ .
\label{infpot}\end{equation}
\emph{The infection potential at day $t$ is just the weekly number of new infections $W_{t+2}$ two days later, multiplied by the unknown constant $c$ which changes slowly with time.} This means that the shape of curves in Figure \ref{fj1}  reflects the course of the epidemic, including asymptomatic cases. The level must be corrected by a factor for each country.  

We did neglect variations in the period and degree of infectivity of a sick person. Our aim was to get a simple formula. Moreover, weekly sums are already in use  and have the advantage of reducing weekly periods in the numbers of daily new infections \cite{OWID, RKI, Bandt}. 

\subparagraph{Daily and total reproduction number.} 
Once we defined the infection potential, we can study its performance. We assume that the incubation time of Covid-19 is exactly 5 days \cite{worldo}. That is, the infection potential for infections at symptom day $t$ must be taken at $t-5.$ Using equations \eqref{ccc} and \eqref{infpot}, the \emph{daily reproduction number} for infections at symptoms day $t$ is  
\begin{equation} 
r_t=\frac{\mbox{infections at day } t}{\mbox{infection potential at day }t-5}
=\frac{I_t}{cW_{t-5+2}}= \frac{N_t}{W_{t-3}} \ .
\label{daily}\end{equation}
The essential point is that the unknown constant $c$ cancels out.
Moreover, since the infection potential is a weekly sum of infection numbers $I_t,$ the total reproduction rate of the infection potential at time $t$ is a sum of seven successive daily reproduction rates:
\begin{equation} 
R_t=r_t+r_{t-1}+...+r_{t-6} \ .
\label{total}\end{equation}
This is written as a backward sum since it will be calculated daily for the most recent time point $t.$ This is the `observed reproduction rate' at observation time $t.$ It refers to the infections which took place 2-3 weeks ago. For Germany, we would date it back by 17 days - 14 days delay between infection and observation plus 3 days because $r_{t-3}$ is the middle term of \eqref{total}. However, an exact estimate of this delay is only necessary when we investigate the effect of certain lockdown measures. The main purpose of a reproduction rate is the study of the present and the prediction of observations of the next few days.

\subparagraph{The smooth reproduction rate $R^*$.} 
Note that \eqref{total} is again a moving sum with a smoothing effect. With length 7, it will further reduce weekly periods in the data.  Nevertheless, the graph of $R_t$ often has plenty of little corners in case of irregularly collected data $N_t.$ For this reason, we use a smooth version of daily reproduction rate. In the formula $r_t= N_t/ W_{t-3}$ we replace the numerator by $\frac12 (N_t+N_{t-1})$ and the denominator by $\frac13 (W_{t-2}+W_{t-3}+W_{t-4}).$ This accounts for the fact that the incubation period is not constant, and not all cases have been tested on their symptoms day. The resulting daily and total reproduction rates are
\begin{equation} 
r^*_t=\frac{3(N_t+N_{t-1})}{2(W_{t-2}+W_{t-3}+W_{t-4})}\quad
\mbox{ and }\quad R^*_t=r^*_t+r^*_{t-1}+...+r^*_{t-6} \ .
\label{sta}\end{equation}
This smooth version was used for Figure \ref{fj1}. As we see below, there is little difference between $R$ and $R^*.$
Since the most recent value $N_t$ is used only with factor $\frac12 ,$ we lose half a day in actuality, but we gain a lot of smoothness. We tried a number of similar smooth versions, and found that $R^*$ is a good choice although random variations in incubation period, period of being infectious, day of testing minus symptoms day etc. are larger than expressed by \eqref{sta}. A longer moving sum in numerator or denominator would mean further loss of actuality, however.

\begin{figure}[h!t] 
\begin{center}
\includegraphics[width=0.49\textwidth]{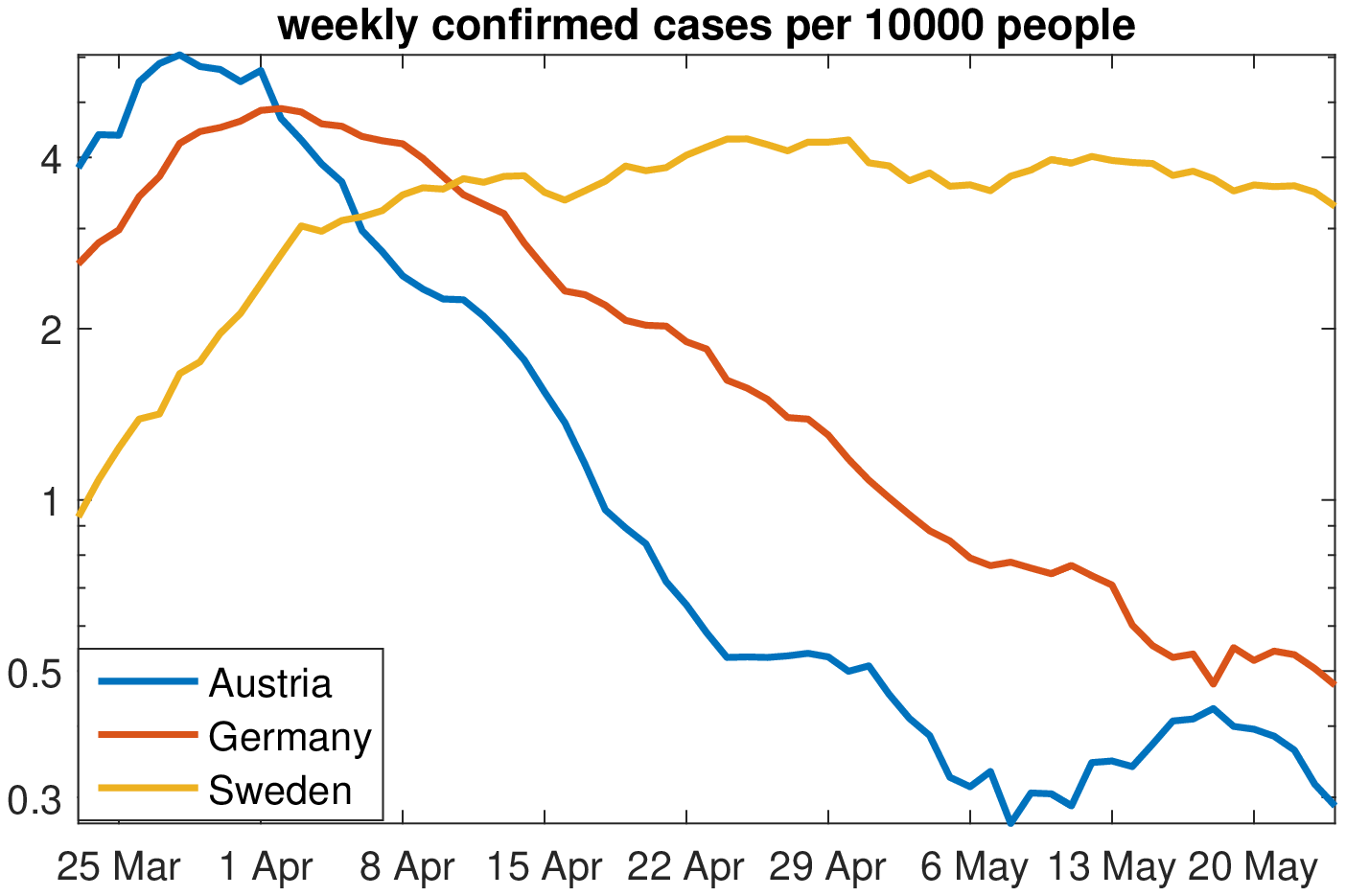}\ 
\includegraphics[width=0.49\textwidth]{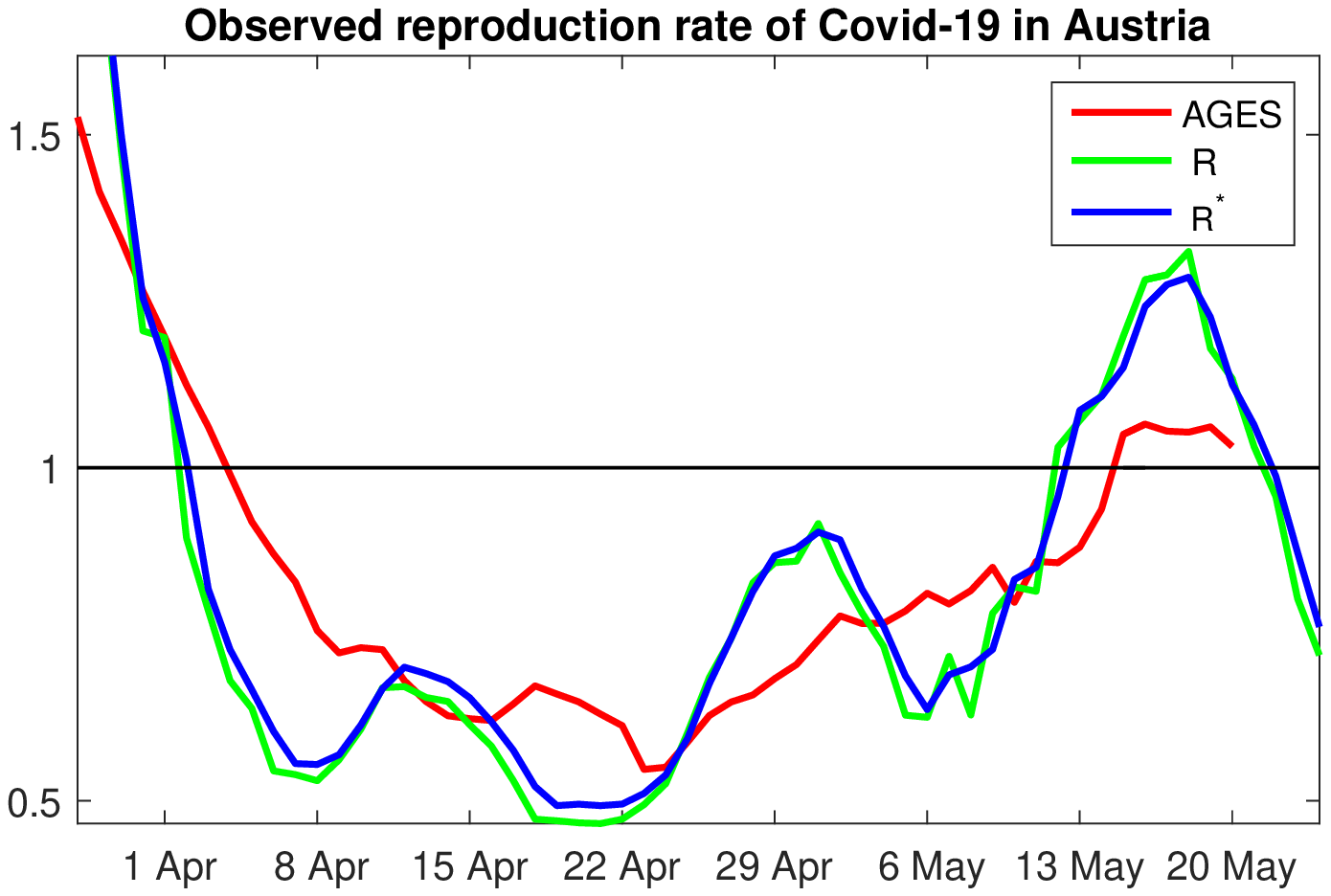} \\
\includegraphics[width=0.49\textwidth]{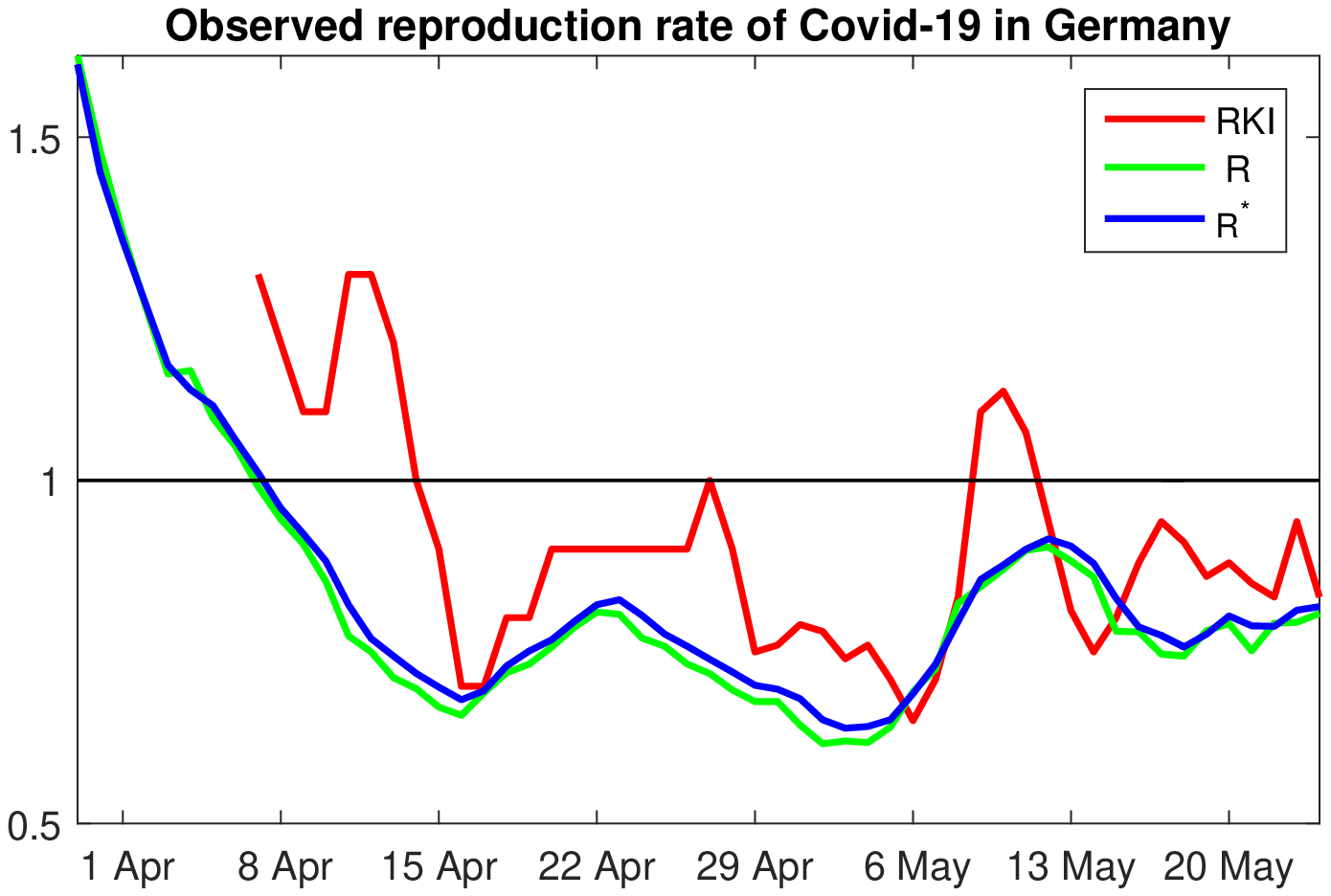} \
\includegraphics[width=0.49\textwidth]{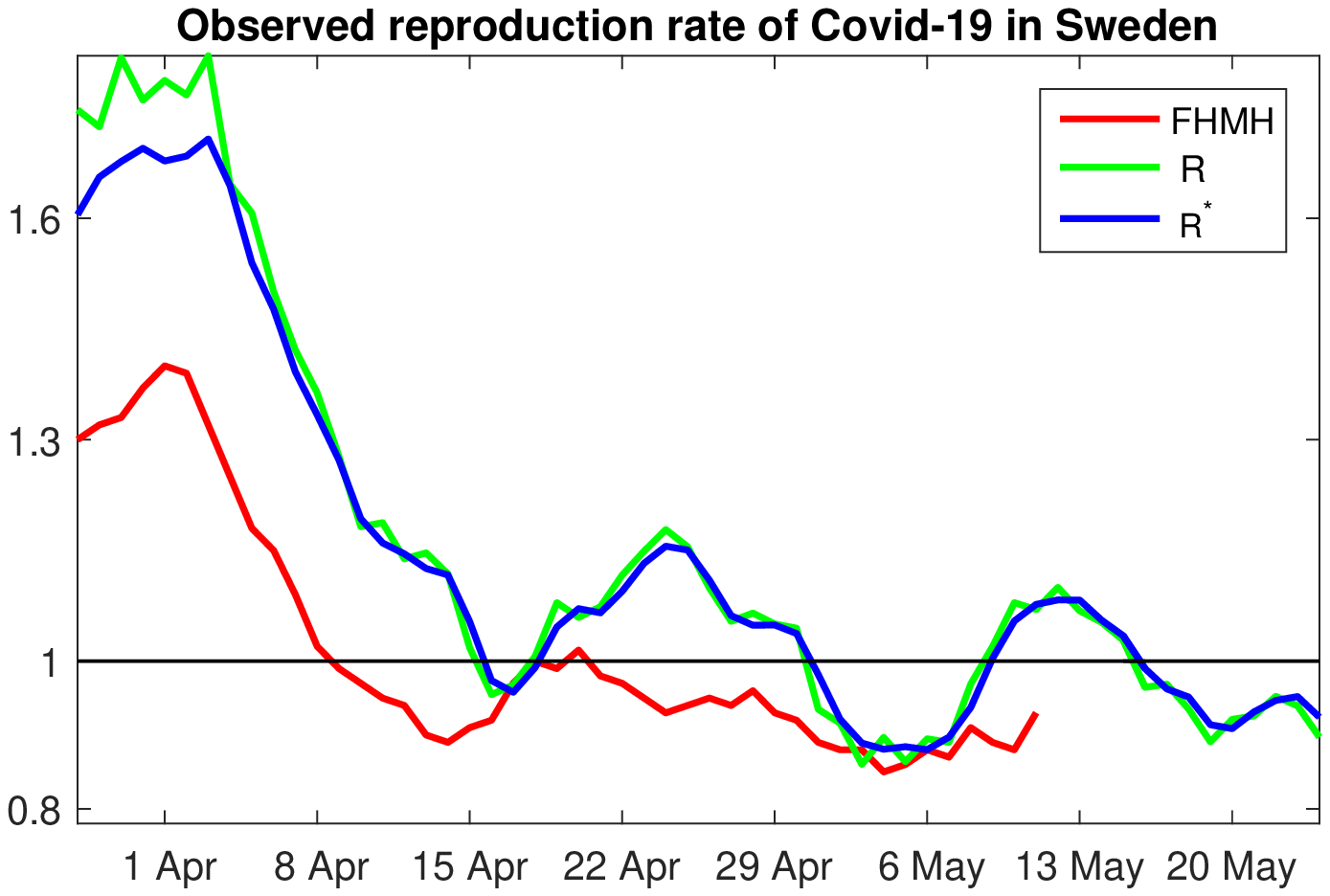}  
\end{center}
\caption{Weekly incidences for Austria, Germany, and Sweden, March 25 to May 24, 2020, and official reproduction functions for each of the countries, compared with our $R$ and $R^*.$}\label{fj3}
\end{figure}

\subparagraph{Comparison with established methods.} 
Let us compare $R$ and $R^*$ with the reproduction numbers used by health authorities in Austria, Germany, and Sweden.  Figure \ref{fj3} shows the weekly incidences for the three countries. The level of Germany and Austria is comparable. Sweden did test only severe cases and consequently has a higher incidence. Sweden has been in the center of public interest since the pandemic is managed without hard lockdown. Because of the multiplicative nature of reproduction, incidences are presented on logarithmic scale. 

The health authority AGES in Austria calculates reproduction rates by a method of Cori et al. \cite{Cori} which beside case numbers requires the distribution of serial intervals - time distances between a positive case and a person infected by that case. They estimated the distribution using their contact tracing work. Details are found in \cite{ages}, reproduction numbers can be downloaded from the website. The method works for small regions, even for very few cases.  All cases are dated to the time of the first positive laboratory test. The reproduction function is slowly varying. The functions $R_t$ and $R^*_t$ show more variation, indicate the changes between $R>1$ and $R<1$ two days earlier than the reproduction function of AGES, and are more sensitive with respect to the recent increase of case numbers on low level.

In Germany, the Robert Koch Institut is the responsible authority for the epidemic. Reproduction rates are given in the daily reports \cite{RKI}, the method is explained in \cite{epibul,RKR}. Using a probabilistic method, cases are dated back to the onset of the sickness. Afterwards, $R$ is calculated as a quotient of two weekly sums with difference of four days, which is assumed to be the serial interval. This method is a bit unstable and resulted in values $R\ge 1$ during a time when weekly case numbers decreased continuously. The important change from increase to decline of cases was confirmed by our numbers several days earlier than by the official ones \cite{Bandt}.
In this comparison, we used the data of \cite{RKI} for determining $R$ and $R^*.$

In Sweden, the health authority uses the method of \cite{Cori} as in Austria, but with a serial interval estimation of Nishiura et al. \cite{Nishi} who found a mean value of 4.8.  Their reproduction function went below 1 before April 10 and stayed below one. Compared with the weekly averages in Figure \ref{fj3} this seems too optimistic. Our $R$ and $R^*$ are clearly larger in March, and then fluctuate around one. We think this describes the development of Swedish case numbers more accurately.

Our simple reproduction rates perform well also for American data, where a Bayesian methodology was used to compute $R$ on \url{rt.live}, and for Brazil, where Figure \ref{fj1} confirms the predictions of \cite{rep21}. 
There is always little difference between $R$ and $R^*.$ The smooth version looks nicer and seems more convenient for prediction. 

\bibliographystyle{plain} 
\bibliography{copre2}

\end{document}